# Vectorial dispersive shock waves on an incoherent landscape

J. NUÑO[1,2], C. FINOT[1,]*  AND J. FATOME[1,3]

[1]*Laboratoire Interdisciplinaire Carnot de Bourgogne, UMR 6303 CNRS - Université Bourgogne-Franche-Comté, 9 av. A. Savary, Dijon, France*
[2] *Departamento de Electrónica, Universidad de Alcalá, 28805 Alcalá de Henares (Madrid), Spain*
[3] *Department of Physics, The University of Auckland, Private Bag 92019, Auckland 1142, New Zealand*
*\*Corresponding author: christophe.finot@u-bourgogne.fr*

**We study numerically and experimentally the impact of temporal randomness on the formation of analogue optical blast-waves in nonlinear fiber optics. The principle-of-operation is based on a two-components nonlinear interaction occurring between a partially coherent probe wave co-propagating in a normally dispersive optical fiber together with an orthogonally polarized intense short pulse. The cross-polarized interaction induces a dual phase-singularity in the probe profile which leads to the formation of two sharp fronts of opposite velocities. An optical blast-wave is then generated and leads to an expanding rarefication area surrounded by two dispersive shock waves which regularize the shock onto the probe landscape. Here we focus our study on the impact of randomness in the shock formation. In particular, we show that the lack of coherence into the probe wave acts as a strong diffusive term, which is able to hamper or inhibit the shock formation. Our experimental observations are confirmed by numerical predictions based on a system of two incoherently coupled nonlinear Schrödinger Manakov equations.**

***OCIS codes:*** *(190.0190) Nonlinear optics; (060.4370) Nonlinear optics, fibers; (060.5530) Pulse propagation and temporal solitons.*

## 1. INTRODUCTION

The regularization of gradient-catastrophe phenomena through dispersive shock waves (DSWs) constitutes a universal mechanism experienced in numerous fields of science, ranging from hydrodynamics, atmospherics, quantum superfluids or nonlinear optics. Direct manifestations of DSWs can be also encountered in nature with the emergence of mascaret waves along river estuaries or the formation of atmospheric morning glory roll clouds [1-3]. Fundamentally, DSWs phenomena occur in conservative, low-dissipative or inviscid media and is triggered by means of two requisite ingredients: nonlinearity of the medium and wave dispersion. The most common scenario of DSWs generation consists in the regularization of a step-like gradient catastrophe through dispersion, which leads to the emergence and expansion of a non-stationary wave fan structure connecting the lower and upper states of the step [2, 3]. Nonlinear optics, and especially optical fibers, have been proved to be a powerful and well suited environment to mimic and investigate the complex dynamics of such wave structures [3-12]. Indeed, it is now well-accepted that in weakly dissipative nonlinear Kerr medium, under the assumption that nonlinearity dominates chromatic dispersion, at least in the early stages of propagation, the nonlinear Schrödinger equation that describes the field evolution in the fiber can then be reduced to shallow water wave equations [7-11]. In such configuration, any incident light modulation will experience a strong temporal steepening, leading to the formation of an intensity gradient catastrophe, followed by a dispersive regularization through DSWs emission. Thanks to the reliability, maturity of available equipment's and robustness of fiber systems, numerous experimental demonstrations have been proposed to trigger, generate and study the complex dynamics of these DSWs [5-12]. However, most of the research effort has been focused so far on experimental configurations restricted to the scalar case, thus involving a single mode of propagation, with pioneering experiments reported by Rothenberg and coworkers in 89 [4]. Very recently, we have proposed to extend DSWs phenomena to the vectorial dimension offered by the polarization of light propagating in optical fibers. More precisely, by means of a specifically engineered two-component vectorial system, we have successfully demonstrated the generation of blast-waves occurring in optical fibers under the action of a double optical piston [12]. This experimental demonstration relies on a vectorial interaction through cross-phase modulation (XPM) between a weak-continuous wave (CW) probe, co-propagating in a normally dispersive optical fiber together with an orthogonally polarized intense short pulse. In such a configuration, the strong defocusing regime of the fiber first induces a sharp steepening of the pulsed signal. Subsequently, the increasing sharpness and temporal expansion of the pulse edges then induce a gradient-catastrophe chirp profile on the orthogonally polarized CW probe through the incoherent XPM mode coupling. Combination with normal dispersion leads to the creation of two fronts of opposite velocities, thus blasting the energy from a central gap in the CW probe profile, surrounded by two fast oscillating DSWs, similarly to an optical axe [13, 14]. Experimental demonstration was found in very good agreement with numerical predictions provided by a vectorial system of two coupled Manakov equations [12].

In this complementary contribution, we further explore the nonlinear interaction occurring between the intense pulsed signal and the orthogonally polarized weak probe, through the generation of DSWs in presence of disorder, introduced here with partial incoherence in the probe landscape. The paper is organized as follows. After first recalling for pedagogical purpose the action of the double-piston induced optical axe on a fully coherent probe wave, we then report novel results by evaluating the impact of randomness into the probe landscape, for which strong and fast temporal fluctuations act as a diffusive term for the DSWs formation. Our numerical predictions are confirmed by a set of experiments for which the level of diffusive disorder contained in the probe wave is found to strongly counteract, damp and even inhibit the shock formation.

## 2. MODELING AND PRINCIPLE OF OPERATION

### A. Modeling

The system under study consists of a cross-polarized interaction between a weak CW probe and an orthogonally polarized intense short pulse co-propagating in a normally dispersive optical fiber. In this work, both waves are characterized by the same central frequency, thus propagate at the same group-velocity. In the framework of nonlinear fiber optics [15], the evolution of the complex slowly varying amplitudes of the intense pulse $u$ and the probe $v$ can be described by the following set of two incoherently coupled nonlinear Schrödinger (NLS) equations for which, averaging the nonlinear contribution over fast and random polarization motions in km-long fibers leads to the well-known Manakov system [16, 17]:

$$\begin{cases} i\dfrac{\partial u}{\partial z} + \dfrac{\beta_2}{2}\dfrac{\partial^2 u}{\partial t^2} + \dfrac{8}{9}\gamma\left(|u|^2 + |v|^2\right)u + i\dfrac{\alpha}{2}u = 0, \\ i\dfrac{\partial v}{\partial z} + \dfrac{\beta_2}{2}\dfrac{\partial^2 v}{\partial t^2} + \dfrac{8}{9}\gamma\left(|u|^2 + |v|^2\right)v + i\dfrac{\alpha}{2}v = 0 \end{cases} \quad (1)$$

where, $z$ and $t$ represent the propagation distance and time coordinate in the comoving frame of the pulse. $\gamma$ corresponds to the nonlinear Kerr coefficient and $\beta_2$ the group velocity dispersion of the fiber, whilst $\alpha$ indicates propagation losses. The factor 8/9 is applied to the Kerr coefficient in the Manakov model so as to take into account for polarization randomness along the fiber length [17] In the following, for convenience, we may call the pulse $u$, the pump even if no transfer of energy occurs between $u$ and $v$. Indeed, it is important to note that no transfer of energy occurs between the two waves during the nonlinear cross-polarized interaction. In fact, as shown by the nonlinear XPM terms in Eqs. (1), both waves are incoherently coupled in every $z$ by a phase term proportional to the temporal intensity profiles of the waves so that the energy initially contained into the probe remains unchanged (except for propagation losses). Furthermore, in the following, we neglect high-order linear and nonlinear effects such as third- and fourth-order dispersions, self-steepening and Raman scattering which play a minor role in the present experiment.

In order to illustrate the formation and evolution of the shock wave in this vectorial configuration, we consider the following set of parameters which corresponds to the experiment described in section III. The pump $u$ is a super-Gaussian-like chirp-free pulse (second order super-Gaussian) characterized by a full width at half-maximum (FWHM) duration of 68 ps, a peak power $P_c$ = 3.8 W and a central wavelength of 1550 nm. The fully-coherent CW probe $v$ is emitted at the same wavelength and is characterized by an average power of 10 mW. The optical fiber involved in our experiment consists in a 13-km long standard dispersion compensating fiber (DCF), typically exploited for dispersion management in telecom transmission systems. Our DCF fiber is characterized by a chromatic dispersion $\beta_2$ = 166 ps$^2$/km ($D$ = –130 ps/nm/km) at 1550 nm, a loss parameter $\alpha$ = 0.4 dB/km and a nonlinear Kerr coefficient $\gamma$ = 6.2 /W/km. It is important to notice that these experimental parameters place our problem in the weakly dispersive regime of NLS (at least in the early stage of propagation). Therefore, this work fully addresses the analogy between nonlinear optics and shallow water hydrodynamics in the first stage of propagation.

### B. Case of a fully coherent continuous probe

In this section and for sake of clarity, we first recall the case for which the weak initial probe $v(t)$ corresponds to a fully coherent CW [12]. The typical results, involving a significant higher power than those reported in [12], are summarized in Fig. 1 thanks to numerical resolution of Eqs. (1). First-of-all, the pump pulse undergoes a nonlinear dynamic typical of the evolution of a high-power short pulse in a normally dispersive medium. More precisely, due to the interplay between self-phase modulation (SPM) and chromatic dispersion, the pump $u$ experiences a significative temporal and spectral expansion (x31 and x15 expansion factor, respectively) and is reshaped into first a broad sharp pulse and then parabolic profile (Fig. 1(a1)) [4, 9, 18, 19] associated with a wide parabolic-like spectrum (Fig. 1(b)) typical of a nonlinear dispersive similariton or spectron [20-22]. Subsequently, the temporal expansion of the pulsed pump follows the dispersive mapping imposed by the fiber and induced by the frequency dependence of the wave velocities due to chromatic dispersion. Consequently, the pulse acquires a close-to-linear chirp profile with a slope governed by $1 / (\beta_2 z)$ (Fig. 1(c)) [20]. Note that such a process of dispersive mapping is quite similar to the technique of dispersive Fourier transform often used for real-time spectral characterizations [23]. Accordingly, the temporal intensity profile of the pump wave then becomes a scaled replica of its parabolic spectral intensity profile [20, 24]. It also appears from Fig. 1(a1) that the cross-phase modulation induced by the orthogonally polarized weak probe does not significantly modify the overall shape of the broadened pulse when compared to its output profile in absence of the CW probe (black dotted line). The temporal consequence of the probe-induced XPM is the emergence of small ripples at the top of the pump profile. In fact, for large propagation distances, due to the ns broadening experienced by the pulsed pump, its peak power has significantly decreased and becomes comparable to the probe one, which in turn shows strong temporal oscillating structures due to DSWs.

In parallel, due to the incoherent XPM coupling, the CW probe $v(t)$ acquires progressively a nonlinear phase shift proportional to the intensity profile of the pulsed pump [25]. The resulting temporal chirp that appears on the continuous wave probe is then mainly characterized by a redshift on the leading part of the probe and a blueshift throughout its trailing edge, leading to a large spectral broadening. As the pump pulse expands and develops steeper and steeper edges, a piston effect is then imposed into the phase profile of the probe, which triggered by chromatic dispersion leads to the creation of two fronts of opposite velocities blasting the energy from the central region. Subsequently, the chromatic dispersion becomes predominant along these vertical fronts, causing the emergence of fast non-stationary oscillations surrounding a temporal gap, thus creating the characteristic imprint of a blast-wave: the optical axe.

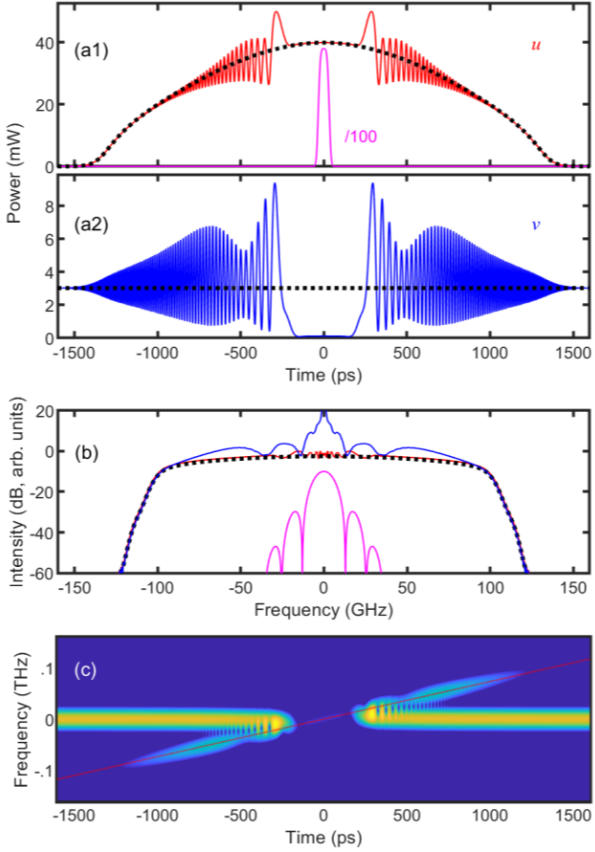

Fig. 1. Temporal and spectral properties of the pump and probe waves when an initial fully coherent CW probe is used. Numerical results of Eq. (1) obtained after propagation in a 13-km fiber. (a) Temporal and (b) spectral intensity profiles of the pump and probe waves. Results in presence of XPM (red and blue curves) are compared with the results achieved when the pump wave propagates solely in the fiber (black dotted lines). The properties of the initial pump wave are also plotted with purple line (/100). (c) Spectro-temporal plot of the probe wave. The red dashed line indicates a slope of $1/\beta_2 z$ that is typical of the far-field propagation in a purely dispersive medium. The colormap is logarithmic and spans on a 20-dB dynamics.

Quite importantly, as long as the SPM of the probe can be neglected, the nonlinear chirp experienced by the CW probe through the XPM nonlinear coupling is identical to the SPM undergone by the pulsed pump. Consequently, as displayed in Figs. 1(a) and (b), the pump and probe waves are characterized by both similar temporal and spectral expansions. This piston effect induced DSWs is also readily visible when using a time-frequency representation [26, 27]. More precisely, the spectrogram depicted in Fig. 1(c) clearly highlights the formation of a temporal gap in the central part of the CW probe. Furthermore, located along both edges of this rarefaction area, a DSW emerges from the fast oscillations occurring from the superposition of the remaining CW probe background and new generated frequencies [9]. The spectrogram also confirms that the frequencies are distributed along a dispersive mapping line characterized by a slope of $1/(\beta_2 z)$. Note that this rarefaction wave, obtained in a single fiber with distributed linear and nonlinear effects, can be in some extend compared with the temporal cloaking process relying on two stages : an XPM-induced spectral broadening in a first highly-nonlinear waveguide followed by purely dispersive propagation [28].

Let us note that contrary to previous works investigating the development of fan structures induced by the nonlinear evolution of a phase-modulated CW [8, 11, 29] many differences are worth to be stressed in our vector system. First of all, in the previous works dealing with scalar propagation, the phase modulation was the initial fully-controlled condition of the CW whereas it is here continuously induced along the propagation distance by the piston pump in a non-trivial manner. Then, the self-phase modulation of the flaticon was a key ingredient of the spectro-temporal dynamics [8]. In the present work, SPM of the probe signal can be neglected and the nonlinear phase modulation is induced not by the wave itself but by the orthogonally polarized co-propagating pulse. The evolution of the DSWs is therefore intrinsically coupled to the evolution of the pump wave. Similar differences holds when comparing our configuration with the recently published pattern resulting from the nonlinear propagation of a dark perturbation in a nonlinearly focusing medium where, in this case, modulation instability plays a crucial role [30]. Finally, since the present blasting DSWs develop onto a continuous wave background, the resulting pattern appears very different from the features numerically described in XPM-induced optical wave-breaking [31].

**C. Case of a partially coherent probe landscape**

Compared with our previous contribution dealing with the fully coherent case [12], here we complete the discussion by focusing on the impact of randomness on the generation of vectorial DSWs. In analogy with mitigation of a ballistic shock wave in a disordered medium, here we study the evolution of optical blast waves in a partially incoherent landscape. Some examples of the impact of randomness in coupled fiber optics systems have already been studied in the past, but the coupling mechanism was essentially driven by a gain process [32, 33]. On the contrary, our configuration does not imply any transfer or increase of energy. Furthermore, in contrast to some of our previous works involving nonlinear evolution of partially coherent waves in vectorial fiber systems [34, 35], the two waves are here characterized by very different levels of power and one of the two waves, i.e. the piston pump, remains fully coherent.

We have carried out a set of numerical simulations based on coupled Eqs. (1). The initial incoherent wave can be modelled in the spectral domain by a Gaussian intensity profile of FWHM width $\Delta f$ with a delta-correlated random spectral phase $\varphi(f)$ uniformly distributed between $-\pi$ and $\pi$:

$$u(f) \propto \exp\left(-2\ln(2)\frac{f^2}{\Delta f^2}\right)\exp\left(i\,\varphi(f)\right). \qquad (2)$$

Results after propagation in the fiber are displayed in Figs. 2 and are averaged over 5000 shots, which enables us to gain a better dynamic on the calculus of the probability distribution function (pdf). We confirm that in the case of low incoherence ($\Delta f$ = 10 GHz), the main features of the temporal structure of the vectorial DSWs as shown in Fig. 1 for the coherent case are qualitatively reproduced (see panels a). A temporal gap gets clearly opened and its duration is similar to the result achieved in the coherent case. It should be noted that averaging the results obtained over time (blue line) does not exactly reproduce the results achieved for a fully coherent probe (white line), the amplitude of the fast oscillations observed in the pump and probe profiles being significantly damped by the presence of randomness.

Much higher discrepancy is visible when the spectral width of the partially coherent landscape becomes of the same order as the XPM-induced frequency shifts imprinted by the piston pump. In this case, the central rarefaction zone tends to almost vanish (panel (b1) obtained for $\Delta f$ = 40 GHz), whereas a strong damping of the DSWs fan structure can be clearly observed. Using numerical simulations, we increased further the amount of disorder and tested the case for which the level of incoherence is now comparable with the highest frequency shifts induced by XPM. Results for a 100-GHz initial incoherence is plotted in panel (b2) for which the vectorial DSW has been nearly fully blurred. The intensity profile averaged over 5000 simulations is not perfectly continuous but presents a slight decrease in the central part surrounded by extremely slight bumps (see the fluctuations magnified by a factor 8, white line panel (b1)).

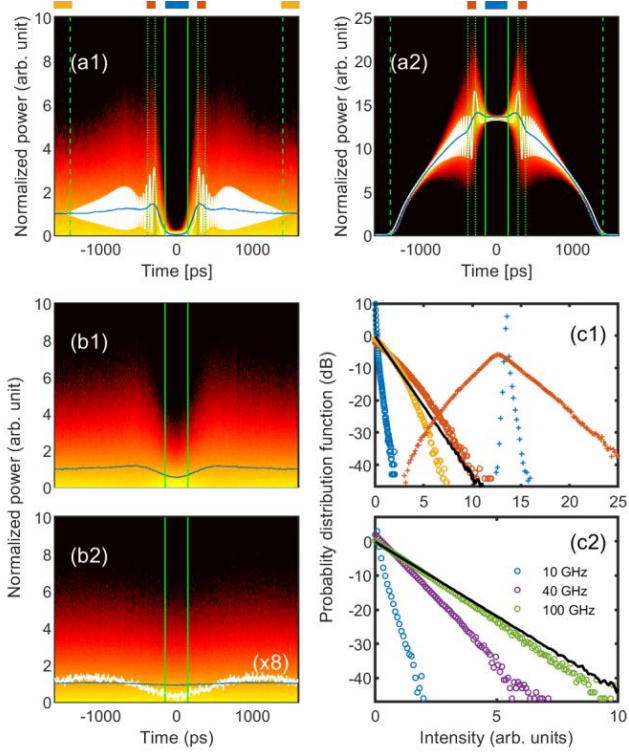

Fig. 2. Impact of the partial coherence on the probe signal. Results obtained from numerical simulations of the temporal intensity profile. (a) Output landscape for a spectral incoherence of 10 GHz. Results for the probe and pump are plotted on panels (a1) and (a2) respectively. The white line corresponds to the results achieved for a CW probe. (b) Influence of the coherence of the probe: results obtained for a coherence of 40 and 100 GHz are plotted on panels (b1) and (b2) respectively. The white line of panel b2 corresponds to the fluctuations magnified by a factor 8. The blue solid lines in panels (a) and (b) are for the average behavior. (c) Evolution of the pdf of the pump (cross) and probe (open circles). Panel (c1) summarizes the properties in various parts of the waves for a spectral incoherence of 10 GHz. Pdf of the central part (located between the continuous green lines in panel a2) are plotted in blue whereas properties in the lateral fluctuating zone (between dotted-lines in panels (a)) are plotted in red. Properties of the probe are plotted with yellow circles for the region outside the temporal pump expansion (outside the dashed zone plotted in panels (a)). Panel (c2): statistical properties of the probe in the central region for different levels of incoherence: results obtained for 10 GHz, 40 GHz and 100 GHz are reported with blue, purple and green circles respectively. In panels (c), the continuous black line represents the input probe properties (corrected from the linear losses). Results are averaged over 5000 shots.

The statistical distributions of the temporal intensity profiles are reported in panel (c1) for a 10-GHz incoherence. In the region of the pump humps (red area), as the dispersion translates the XPM-induced incoherent phase into power fluctuations, the pump exhibits a broad distribution of the peak-power (with a ratio of the mean value $M$ with respect to the standard deviation $\sigma$ of $M/\sigma$ = 5.8). On the contrary, the pump is unaffected close to its maximum (blue area, $M/\sigma$ > 100). The pdf of the probe is strongly depleted in the central part, highlighting the nearly complete depletion of this area even in presence of an initial incoherent landscape. The part of the pulse that does not interact with the pump (yellow circles) is less tailed than the initial distribution (characterized by a linearly decreasing pdf when plotted on logarithmic scale [36]), in agreement with the evolution in the regime of normal dispersion reported for turbulent waves [37]. On the contrary, the pdf evaluated in the hump regions exhibits higher values than the exponential distribution of the input Gaussian probe field.

Finally, the pdf calculated in the central region of the probe for different levels of initial incoherence (panel (c2)) confirms that increasing the amount of disorder progressively inhibits the rarefaction area opened by the optical axe so that the pdf gets closer and closer from the input pdf. We attribute this behavior to the fact that increasing the level of randomness into the probe landscape enhances the diffusion of the medium through chromatic dispersion and thus invalidates the shallow water approximation. Indeed, the temporal random fluctuations included in the probe profile, typically much faster than the pulse duration, are highly diffusing and thus hamper the shock formation. These results can be also easily understood by analogy with a disordered material or a discrete medium for which randomness-induced scattering will absorb the shock, just like previously observed in an aqueous dye solution of silica spheres pumped by a Gaussian laser beam [6].

We plotted in Fig. 3(a) the optical spectra of the probe that can be simulated by averaging the 5000 numerical shots, before (dashed-lines) and after propagation in the nonlinear optical fiber (solid lines). For the 10-GHz incoherence case (in blue), we can make out that the spectral broadening experienced by the probe is much larger than the initial incoherence. The resulting spectrum is very close to the one achieved for a purely coherent CW probe (black solid line). For a 40-GHz linewidth of incoherence (in red), the impact of the cross-phase modulation only affects the wings of the pulse. For higher levels of incoherence, 100 GHz (in yellow), the changes in the spectra become nearly invisible. Finally, we also plotted in Figs. 3 various single shot spectrograms that can be achieved for the different levels of incoherence under study. For 10 GHz (panels b), one may retrieve from each individual spectrogram the features already observed for the fully coherent case reported in Fig. 1(c). Indeed, the main features are fully observed when averaging over 10 shots: a gap is clearly opened in the central part and the frequencies are distributed along a line with a slope of $1/\beta_2 z$. Increasing further the incoherence to $\Delta f$ = 40 GHz (panels c), one can only recognize, on some realizations, the properties of the vectorial DSWs. When averaging over 20 shots, the overall picture becomes clearer even if the central area is now partly filled. Finally, when the initial incoherence is increased above the spectral extent of the broadened pump wave ($\Delta f$ = 100 GHz, panels d), any signature of the vectorial DSWs can be then retrieved on the single-shot spectrograms. A blurred shadow of the shock may be guessed when averaging is carried out over 100 of realizations (panel d4).

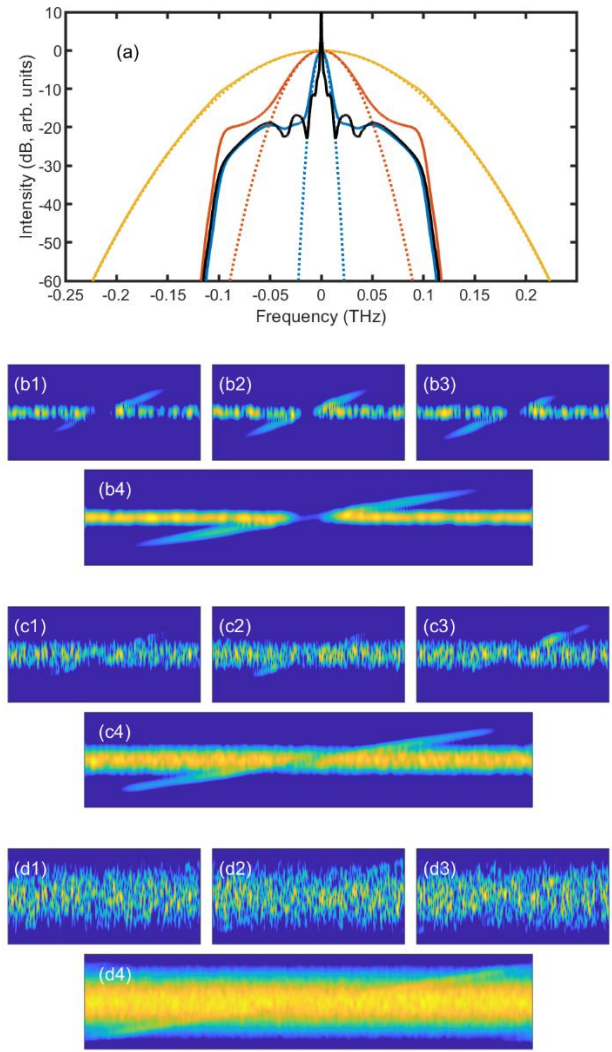

Fig. 3. (a) Optical spectra before and after nonlinear propagation (dotted and solid lines respectively). Results achieved with an initial incoherence of 10 GHz, 40 GHz and 100 GHz are plotted with blue, red and yellow curves respectively, and are compared with the results achieved with a CW probe (solid black line). (b-d) Spectro-temporal plots of the probe wave. Panels b, c, d are for an initial incoherence of 10 GHz, 40 GHz and 100 GHz, respectively. Panels 1-3 are three random shots whereas panels 4 are the average of 5, 20 and 100 shots for 10 GHz, 40 GHz and 100 GHz, respectively. The horizontal time scale and the vertical frequency scale are identical to panel (c) of Fig. 1.

## 3. EXPERIMENTAL VALIDATION

### A. Experimental setup

In order to study experimentally the impact of randomness in the formation of blasting DSWs, we have implemented the setup displayed in Fig. 4. Our setup mainly consists of commercially available devices widely used in the telecommunication area. Moreover, the all-fibered nature of the setup ensures a strong stability over several hours of operation. First-of-all, 68-ps FHWM pulses cadenced at a repetition rate of 312.5 MHz are carved in a CW laser centered at 1550 nm by means of two Mach-Zehnder modulators driven by an electrical 14-Gbit/s pulse-pattern generator (PPG). Note that two cascaded intensity modulators (IM1 & IM2) have been involved in this setup so as to achieve an extinction ratio higher than 40 dB and thus prevent any deleterious interference occurring in between the pump wave and its residual CW background [9, 22]. This pulse train is then amplified to reach a peak power of 3.8 W thanks to an Erbium-doped fiber amplifier (EDFA) and used as piston pulsed pump.

The probe wave first consists of a 10-mW CW landscape which corresponds to a half-portion of the initial 1550-nm signal to guarantee a perfect velocity matching. Furthermore, to study the influence of the randomness of the medium, here provided by the coherence properties of the probe wave, the CW can be replaced by a partially coherent wave generated from a spontaneous noise emission source (ASE from an Erbium-doped fiber amplifier) that is subsequently spectrally sliced and polarized. The optical bandpass filter under use has a Gaussian shape, is centered at the same frequency than the pump pulse (group velocity matching) and its FWHM $\Delta f$ can be tuned between 10 and 100 GHz, leading to typical temporal fluctuations of the probe intensity profile ranging from 44 to 4.4 ps, respectively. The average power of this partially coherent probe wave is also fixed to 10 mW.

Pump and probe signals are then orthogonally polarized by means of two polarization controllers (PC) and combined thanks to a polarization beam splitter (PBS) before injection into a 13-km long normally dispersive fiber (DCF) whose parameters are listed above in the modeling section. At the output of the fiber, the pump and probe waves are split in polarization thanks to a second PBS and characterized in the spectral domain using an optical spectrum analyzer (OSA). The output temporal profiles of the pump and probe waves are also recorded by means of a 70-GHz bandwidth sampling oscilloscope combined to two high speed 70-GHz photodetectors. A 50-GHz real-time oscilloscope is also used to resolve the shot-to-shot evolution of the partially coherent probe wave.

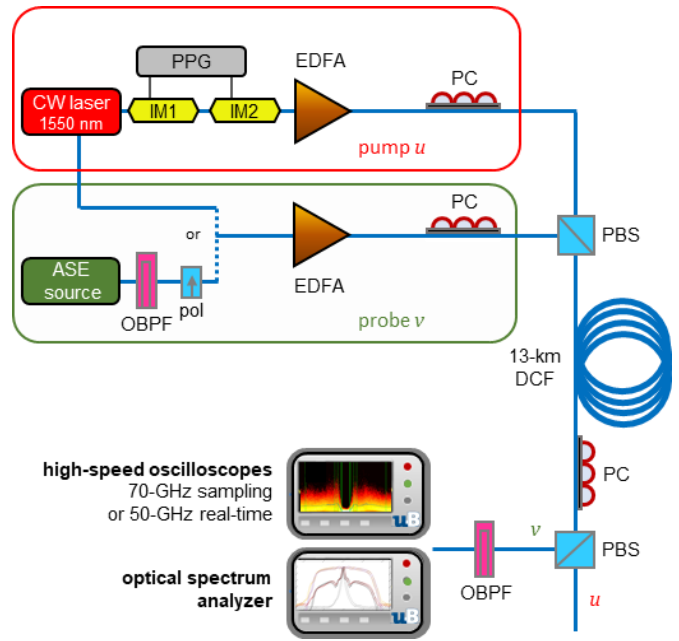

Fig. 4. Experimental setup. CW : Continuous Wave, ASE : Amplified Spontaneous Emission, PPG : Pulse Pattern Generator, IM : Intensity Modulator, EDFA : Erbium Doped Fiber Amplifier, PC : Polarization Controllers, PBS : Polarization beam splitter/combiner, DCF : Dispersion Compensating Fiber, OBPF : Optical BandPass Filter, Pol : Polarizer.

## B. Experimental temporal and spectral profiles obtained for a CW probe

Figure 5 displays the experimental temporal and spectral properties of the pump and CW probe waves recorded at the input and output of the fiber when the peak power of the pump pulse is set to 3.8 W. We can first notice in panel (a1) the significant temporal broadening of the pump pulse for which the FWHM increases from 68 ps in input of the system (pink line) to more than 1.65 ns after propagation in the 13-km DCF (red line). Note that this temporal broadening roughly corresponds to the expansion of the shock fan displayed in panel (a2). Furthermore, the output pulse shape can be well approximated by a parabolic profile, depicted here with black circles, confirming the expected dispersion mapping with respect to long propagation distances [21]. We can also note the presence of small oscillations on the top-edges of the output pump profile when compared with the case without CW probe (black line), thus confirming the slight nonlinear contribution of the vectorial DSWs on the pump wave through the XPM coupling, as expected above in Fig. 1(a1).

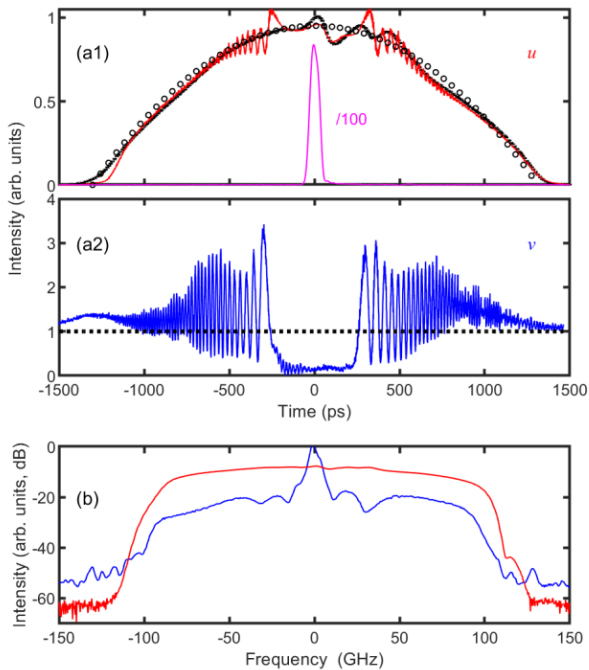

Fig. 5. Experimental results obtained at the output of the fiber for a pump power of 3.8 W. (a1) Temporal profile of the pump wave with (red) and without (black) the co-propagating CW probe. (Circles) Comparison with a parabolic shape. The properties of the initial pump wave are also plotted with pink line (/100). (a2) Temporal profile of the CW probe in input (black dashed-line) and at the output of the fiber (blue). (b) Corresponding output spectral profiles.

Regarding the output probe intensity profile, reported in Fig. 5(a2) with blue solid-line, the piston effect induced by the intense pump wave is readably highlighted. The nonlinear phase shifts induced by the expansion and steepening of the pulse edges create two moving fronts starting from the center and propagating in outwards directions, thus leading to a clear signature of blasting DSWs. Thanks to our 70-GHz bandwidth detector and large cumulated dispersion within the DCF fiber, the oscillating temporal structures of the DSWs can be well resolved. Once again, an excellent qualitative agreement is obtained between our measurements and numerical results of Fig. 1, thus validating the modeling of this vectorial DSW phenomenon by means of the Manakov equations. The slight asymmetry observed in the DSW profile can be attributed to the small asymmetry of the input pulse. These results confirmed the general behavior observed in our previous publication, here obtained for more than twice the pump peak-power involved in our ref. [12], thus enabling a significantly larger temporal expansion of the shock waves.

The nonlinear reshaping of the pump wave is also accompanied by a strong spectral broadening, reported in Fig. 5(b) with red solid-line, characterized by an output total spectral expansion above 200 GHz. Note that the output spectrum of the probe signal (blue solid-line) is also widely broadened with a spectral extension rather similar, evidencing the strong nonlinear coupling that appears between the vectorial dispersive shock wave and the piston pump due to XPM. Note finally that if the temporal and spectral intensity profiles of the pulsed pump are directly linked by the dispersive Fourier transform occurring in the DCF fiber [23], this is clearly not the case for the probe wave due to the presence of the continuous background.

## C. Impact of randomness in the probe landscape

In order to experimentally investigate the impact of randomness in the blasting DSWs induced by a partial incoherence of the probe wave, we then replace the CW probe signal by the filtered ASE source (see setup). Figure 6(a1) shows the results recorded for a 10-GHz temporally incoherent landscape after nonlinear propagation in the DCF fiber in presence of the 3.8-W pulsed pump. This panel displays the accumulation of a thousand of recorded traces. When the pump is switched on, one can clearly observe the typical temporal gap emerging from the fluctuations of the 10 GHz partially coherent wave. We can also make out the existence of bumps localized on each side of the temporal gap, typical of this vectorial DSW dynamics observed above in the case of a CW probe wave. The details of the fan structure are however completely blurred by the incoherence of the probe. Moreover, details of the deterministic pump wave reported in panel (a2) show that the level of fluctuations may strongly vary along the pulse width. Indeed, significant intensity fluctuations surrounding the central part of the pulse are observed, confirming once again the XPM induced by the DSWs around the rarefaction area. In contrast, the central part and the wings of the parabolic profile remain fully coherent. These experimental observations as well as the averaged temporal traces plotted in blue in panels (a1) and (a2) are fully consistent with the numerical conclusions presented in Fig. 2(a).

In order to further assess the statistical behavior of the waves, we have also experimentally evaluated the probability distribution function of the intensity profiles along different temporal areas of the waves. Results are summarized in panels (b1) and (b2) for the probe and pump waves, respectively. Whereas the rarefaction area of the probe (open blue circles, blue region in panel (a)) presents a narrow distribution centered close to zero, the bumps (red full circles, red regions in panel (a) are characterized by temporal events presenting higher peak power than the ones appearing without any interaction with the pump (yellow crosses, yellow region in panel (a)). Regarding the pump, the present experiments confirm that the pdf of the central part (rarefaction area of the probe) is much narrower than the ones affected by the DSWs fluctuations (on the edge of the temporal gap).

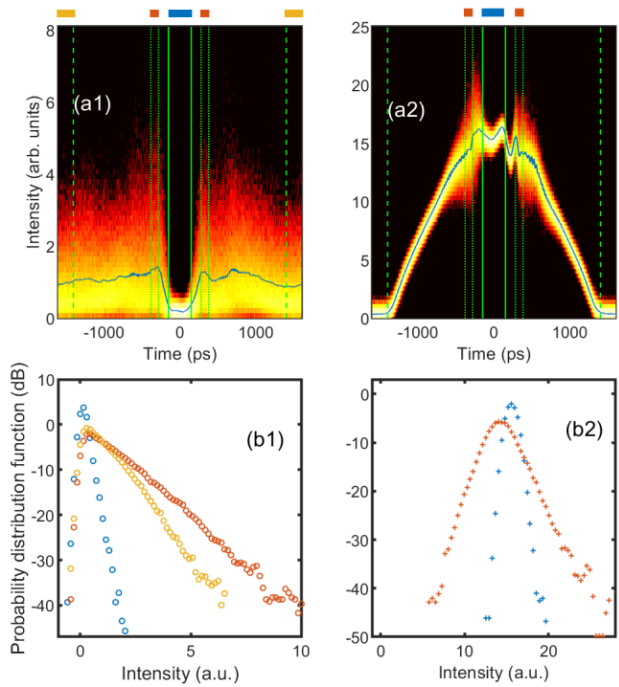

Fig. 6. Impact of the incoherence of the probe wave on the vectorial blasting DSWs. Experimental results obtained for a pump-peak power of 3.8 W and a spectral coherence of the probe of 10 GHz. (a) Output intensity profiles recorded for the probe and pump waves are plotted in panels (a1) and (a2) respectively. (b) Evolution of the pdf in two different temporal sections of the probe and pump waves (panel b1 and b2 respectively). Statistics for the central part of the probe (blue part in panels (a)) are plotted with blue open circles or crosses whereas properties in the lateral bumps (red parts in panels (a)) are plotted in red and in yellow for the region outside the temporal pump extend (yellow part in panels (a)). Intensities are normalized with respect to the average intensity of the probe.

The comparison of the optical spectra obtained in the case of a fully coherent or a 10-GHz partially incoherent probes are shown in Fig. 7(a) and stresses that, as expected from section 1C, the spectral expansion are rather similar and are in both cases imposed by the pump wave. Finally, we investigate in Fig. 7(b) the impact of the level of incoherence in the probe wave by monitoring the temporally averaged DSWs on the sampling oscilloscope triggered by means of the output piston pump for four different levels of randomness. For a 10- or 20-GHz incoherent landscape, we can still observe the piston effect caused by the pump induced XPM on the probe signal and in particular the depletion of the central part. However, in agreement with numerical results presented in Fig. 2, the fast temporal DSW fan structures tend to be damped by the randomness of the probe wave. Indeed, further increasing the amount of disorder tends to even more hamper the shock formation until its total inhibition for an incoherence $\Delta f$ higher than 40 GHz. For an incoherence of 100 GHz, the averaged recording only presents a small bump, consistent with the temporal fluctuations reported in panel (b2) of Fig. 2. Note however that for the 100-GHz incoherence case, the experimental recordings might be limited by the 70-GHz bandwidth of our photoreceiver.

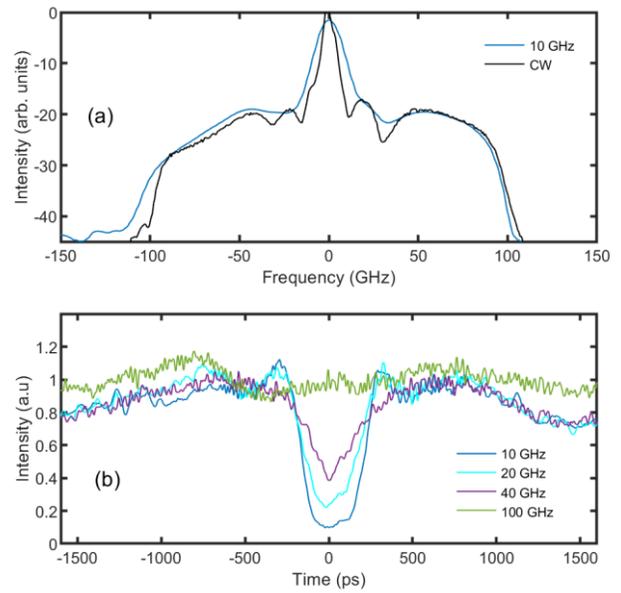

Fig. 7. Impact of the partial incoherence of the probe signal. Experimental spectral and temporal properties of the probe wave recorded at the output of the fiber for a pump-peak power of 3.8 W. (a) Comparison of the output spectrum obtained for an initial fully coherent CW probe (black line) and a 10-GHz partially coherent probe signal (blue line). The resolution of the OSA (0.01 nm) limits the resolution of the initial coherent wave. (b) Output temporal intensity profile of the probe wave after averaging over 1000 shots. Different levels of initial incoherence are compared: 10, 20, 40 and 100 GHz (blue, cyan, purple and green colors, respectively).

## 6. CONCLUSIONS

In this contribution, we have significantly extended our recent results published in ref. [12] and dealing with the study of vectorial double-piston dispersive shock waves. The principle is based on a two-components nonlinear interaction occurring between a weak CW probe co-propagating in a normally dispersive optical fiber together with an orthogonally polarized intense short pulse. This cross-polarized interaction induces gradient-catastrophe phase singularities in the CW probe background which then leads to the formation of two sharp fronts of opposite velocities. An equivalent optical blast-wave is then generated leading to an expanding rarefication area surrounded by two DSWs which regularize the shock onto the initial continuous landscape. Here we have focused our study on the impact of randomness on the shock formation. In analogy with disordered materials or discrete media for which randomness-induced scattering will damp the shock formation, here we have shown that the lack of coherence into the probe wave acts as a strong diffusive term, which is able to hamper or inhibit the shock dynamics. Our numerical predictions are confirmed by a set of experiments in which the level of diffusive disorder in the probe wave is simply tuned by means of its initial spectral width properties. All the experimental results are in full agreement with numerical simulations based on a Manakov system consisting in two coupled nonlinear Schrödinger equations. In conclusion, DSWs are extreme phenomena that are difficult to isolate, study or reproduce in their real environments. Furthermore, the presence and impact of disorder onto dispersive shock wave formation makes this subject even more complex and universal. Therefore, we believe that the present numerical and experimental study fully demonstrate that fiber optics may constitute a fast, convenient and reliable test-bed for DSWs characterization, far beyond the nonlinear optics community. Extension to spatial optics can also be encompassed to benefit from additional degree of freedoms [38].

**Funding Information.** C.F acknowledges the Institut Universitaire de France (IUF). J.N acknowledges the Comunidad de Madrid (SINFOTON2-CM: P2018/NMT-4326 and TALENTO-CM 2017-T2/TIC-5227). J.F acknowledges the financial support from the European Research Council (Grant Agreement 306633, PETAL project) as well as the CNRS and Conseil Régional de Bourgogne Franche-Comté for the mobility program (2019-7-10614).

**Disclosures**. The authors declare that there are no conflicts of interest related to this article.

**Acknowledgment**. All the experiments were performed on the PICASSO platform in ICB. We thank Pr. Miro Erkintalo, Pr. Guy Millot and Dr. Gang Xu for fruitful and stimulating discussions.